\documentclass[aps,prl,twocolumn,showpacs,superscriptaddress,groupedaddress,nofootinbib,preprintnumbers]{revtex4-1}
\pdfoutput=1
\usepackage{amsmath}
\usepackage{amsfonts}
\usepackage{amssymb}
\usepackage{graphicx, rotating}
\usepackage{epstopdf}
\usepackage{epsfig}
\usepackage{latexsym}
\usepackage{multirow}
\usepackage{color}
\usepackage{amsmath,amssymb}
\usepackage{slashed}
\usepackage[colorlinks=true,linkcolor=red,anchorcolor=black,citecolor=blue,filecolor=cyan,menucolor=red,runcolor=filecolor,urlcolor=blue,bookmarks=true,bookmarksnumbered=true]{hyperref}
\definecolor{rossos}{cmyk}{0,1,1,0.55}
\definecolor{bluscuro}{rgb}{0.15, 0.2, .85}
\definecolor{bluchiaro}{cmyk}{1,.3,0.,0.1}

\definecolor{rossos}{cmyk}{0,1,1,0.55}
\definecolor{bluscuro}{rgb}{0.15, 0.2, .85}
\definecolor{bluchiaro}{cmyk}{1,.3,0.,0.1}

\newcommand{\bc}{\begin{center}}
\newcommand{\ec}{\end{center}}
\newcommand{\pMET}{{\bf p}\llap{/\kern1.5pt}_T}
\newcommand{\bea}{\begin{eqnarray}}
\newcommand{\eea}{\end{eqnarray}}
\newcommand{\ignore}[1]{}
\newcommand{\be}{\begin{equation}}
\newcommand{\ee}{\end{equation}}

\def\({\left(}
\def\){\right)}
\def\<{\langle}
\def\>{\rangle}

\def\be{\begin{equation}}
\def\ee{\end{equation}}
\def\bry{\begin{array}}
\def\ery{\end{array}}
\def\bes{\begin{subequations}}
\def\ees{\end{subequations}}
\def\bit{\begin{itemize}}
\def\eit{\end{itemize}}
\def\ben{\begin{enumerate}}
\def\een{\end{enumerate}}

\newcommand{\MET}{E\llap{/\kern1.5pt}_T}
\definecolor{grey}{rgb}{0.6,0.6,0.6}
\definecolor{fuchsia}{rgb}{1,0,1}
\newcommand{\red}[1]{{\color{magenta}\color{red}#1\color{magenta}}}

\begin{document}

\preprint{CERN-PH-TH-2015-019}
\preprint{DFPD-2015/TH/03}

\title{Spot the stop with a $b$-tag}

%

\author{Gabriele Ferretti}
\address{Department of Fundamental Physics, Chalmers University of Technology, 412 96 G\"oteborg, Sweden}
\author{Roberto Franceschini}
\address{Theory Division, Physics Department, CERN, CH-1211 Geneva 23, Switzerland}
\author{Christoffer Petersson}
\address{Department of Fundamental Physics, Chalmers University of Technology, 412 96 G\"oteborg, Sweden}
\address{Physique Th\'eorique et Math\'ematique, Universit\'e Libre de Bruxelles, C.P. 231, 1050 Brussels, Belgium}
\address{International Solvay Institutes, Brussels, Belgium}
\author{Riccardo Torre}
\address{Dipartimento di Fisica e Astronomia, Universit\`a di Padova, and INFN Sezione di Padova, Italy}
\begin{abstract}
The LHC searches for light compressed stop squarks have resulted in considerable bounds in the case where the stop decays to a neutralino and a charm quark. However, in the case where the stop decays to a neutralino, a bottom quark, and two fermions via an off-shell $W$-boson, there is currently a significant unconstrained region in the stop-neutralino mass plane, still allowing for stop masses in the range \mbox{90--140\,GeV.} \,\,In this paper we propose a new monojet-like search for light stops, optimized for the four-body decay mode, in which at least one $b$-tagged jet is required. We show that, already by using the existing 8\,TeV LHC data set, such a search would cover the entire unconstrained region. Moreover, in the process of validating our tools against an ATLAS monojet search, we show that the existing limit can be extended to exclude also stop masses below 100\,GeV.
\end{abstract}
\pacs{12.60.Jv, 14.80.Ly, 13.85.Rm}
\keywords{Supersymmetry, Natural SUSY, Light Stops}

\maketitle

{\emph{Introduction}.}---The top quark gives rise to the leading quantum correction that destabilizes the electroweak scale in the Standard Model (SM). One way to solve this so-called hierarchy problem is to extend the ordinary spacetime symmetries by supersymmetry (SUSY) and introduce new physics at a low scale in the form of superpartners of the SM particles. In order to cancel the leading quantum correction, the superpartner of the top quark, the stop, should have a mass of the order of the electroweak scale and hence be observable at the Large Hadron Collider (LHC). Light stops have been subject to intense recent studies both in the theory \cite{Plehn:2010we,Bornhauser:2010gb,Brust:2011gf,Kats:2011fk,Bi:2011jv,BinHe:2011tw,Drees:2012hw,Bai:2012ux,Plehn:2012mz,Alves:2012gf,Han:2012ve,Kaplan:2012wf,Brust:2012yq,Choudhury:2012hg,Ghosh:2012cx,Evans:2012oq,Kilic:2012yq,Graesser:2012dn,Krizka:2012ii,Franceschini:2012vl,Delgado:2012eu,Dutta:2013gd,Buckley:2013wo,Chakraborty:2013iu,Low:2013kl,Bai:2013wx,Belanger:2013wy,Boughezal:2013cn,Han:2013ui,Dutta:2013wy,Papucci:2014ue,Buckley:2014bf,Czakon:2014wa,Grober:2014vl,Eifert:2014uu,Cho:2014vd} and experimental \cite{Aad:2013ija,Aad:2014qaa,ATLAScollaboration:2014kf,Aad:2014bva,Aad:2014kra,ATLAS:2013aia,CMScollaboration:2013dj,Chatrchyan:2013xna,CMS-PAS-SUS-13-009,CMScollaboration:2013gy,CMScollaboration:2014jo,CMScollaboration:2014tk,CMScollaboration:2014wb,CMS:2014wsa} communities.

Taking a simplified model approach, a key strategy to test R-parity conserving SUSY is to consider only the lightest stop mass eigenstate $\tilde{t}_1$, decaying to the lightest superpartner (LSP), the neutralino $\tilde{\chi}^{0}_{1}$, and taking all other superpartners to be sufficiently heavy and effectively decoupled. In such a simplified model, where the stop mass $m_{\tilde{t}_1}$ and the neutralino mass $m_{\tilde{\chi}^{0}_{1}}$ are the only two parameters, the only relevant SUSY production mode is stop pair production, for which the cross-section is determined by $m_{\tilde{t}_1}$.

In the case where the mass difference between the stop and the neutralino is larger than the top mass, $\Delta m{=}m_ {\tilde{t}_1}\,{-}\,m_{\tilde{\chi}^{0}_1}\,{>}\,m_t $, each of the pair produced stops decays 2-body via an on-shell top, $\tilde{t}_1\,{\to}\, t\,\tilde{\chi}^{0}_{1}$. In this case, for neutralino masses below around 250\,GeV, the current LHC limits exclude stop masses below 600--750\,GeV \cite{CMS:2014wsa,Aad:2014kra}, while for larger neutralino masses there are no bounds.

For mass splittings in the range $m_W\,{+}\,m_b\,{<}\,\Delta m\,{<}\,m_t $, the stop decays 3-body via an off-shell top, $\tilde{t}_1\,{\to}\, b\,W\,\tilde{\chi}^{0}_{1}$, and the stop mass limits reach up to around \mbox{200--300\,GeV}~\cite{Chatrchyan:2013xna,Aad:2014kra,Aad:2014qaa}. It should be noted that the  limits in the different mass splitting regions are not continuously connected
 to each other and close to the mass thresholds at $\Delta m\,{\sim}\,m_t$ and $\Delta m\,{\sim}\,m_W\,{+}\,m_b$, the bounds become weak or disappear.

For even smaller mass splittings, $\Delta m \,{<}\, m_W{+}m_b $, the stop can have two different decay modes, either the \mbox{4-body} decay, via an off-shell $W$, $\tilde{t}_1\,{\to}\, b\,f\,f'\,\tilde{\chi}^{0}_{1}$, or the 2-body decay $\tilde{t}_1\,{\to}\, c\,\tilde{\chi}^{0}_{1}$. While the former decay mode  is phase space suppressed, the latter is 1-loop and Cabibbo-Kobayashi-Maskawa (CKM) suppressed, see e.g.~Refs.~\cite{Delgado:2012eu,Grober:2014vl,Boehm:2000ty,Das:2002zr} for discussions. The branching ratios for these two competing decay modes depend strongly on the flavor structure of the squark soft masses, such as the off-diagonal stop-scharm mixing mass term, as well as the masses of the superpartners that enter the loop. In order to be able to restrict to the minimal set of parameters, it is customary to consider these two decay modes separately, and assume 100\% branching ratio in each case.

In the small mass splitting case, several LHC searches have been optimized for the 2-body charm decay mode. Under the assumption BR${(}\tilde{t}_1\,{\to}\, c\,\tilde{\chi}^{0}_{1}{)}{=}1$, stop masses below 250--300\,GeV have been excluded \cite{ATLAS:2013aia,CMS-PAS-SUS-13-009}. In contrast, no existing LHC analysis has been optimized for the 4-body stop decay mode. Currently, even though the 4-body case is partly covered by the ATLAS searches \cite{Aad:2014kra,ATLAS:2013aia}, there exists a significant unconstrained region in the stop-neutralino mass plane, still allowing for stop masses in the range \mbox{90--140\,GeV} (for neutralino masses below 60\,GeV). Since very light stops could be hiding there, it is important to find means to probe it. In this paper we show that, by augmenting an existing  monojet search with a $b$-tag requirement, this unconstrained region could be covered completely already with the existing 8\,TeV LHC data set.

\begin{table*}[t!]
\small\begin{center}
\begin{tabular}{l|cccccc}
Background				&$t\bar{t}$			&$Z({\to}\nu\nu)$	&$W({\to}\ell\nu)$		&Dibosons		&Others		&Total \\ \hline\hline
M1 (ATLAS~\cite{ATLAS:2013aia})		&$780\pm 73$
	&$~17400\pm 720~$	&$~14100\pm 337~$		&$650\pm 99$
	& $565\pm301$	&$~33450\pm 960$\\ 	
M1+$b$-tag	 			&$307\pm 57\ignore{\red{(35)}}$	&$~261\pm 22$		&$144\pm 7$			&$55\pm 17\ignore{\red{(13)}}$	& -	&$~767\pm 64$\\  \hline
\end{tabular}
\end{center}\vspace{-2mm}
\caption{\small\label{table:BGevents} Estimated numbers of background events with 20.3\,fb${}^{-1}$ of 8\,TeV LHC data. 
}\vspace{-2.7mm}
\end{table*}

{\emph{Existing stop 4-body searches}.}---We start by reviewing the existing searches relevant for stops in the mass range $m_b\,{<}\,\Delta m \,{<}\, m_W{+}m_b$ for which BR${(}\tilde{t}_1\,{\to}\, b\,f\,f'\,\tilde{\chi}^{0}_{1}{)}{=}1$.  In contrast to the case where the stop decays via the 2-body charm decay mode, which has been probed by Tevatron \cite{Aaltonen:2012tq}, CMS \cite{CMS-PAS-SUS-13-009} and \mbox{ATLAS} \cite{ATLAS:2013aia}, neither Tevatron nor CMS have performed any search that places a bound in the 4-body decay case.\footnote{Searches for the stop decay $\tilde{t}_1\,{\to}\,b\,\ell\,\tilde{\nu}$ might have some sensitivity to the 4-body decay mode that we study. However, since the results are not presented in terms of the  4-body stop decay, we do not include them in our summary of the existing bounds.} Therefore we will focus our discussion on two searches performed by ATLAS.

\begin{figure*}[t!]
\begin{center}
\hspace{-4.3mm} \includegraphics[scale=0.37]{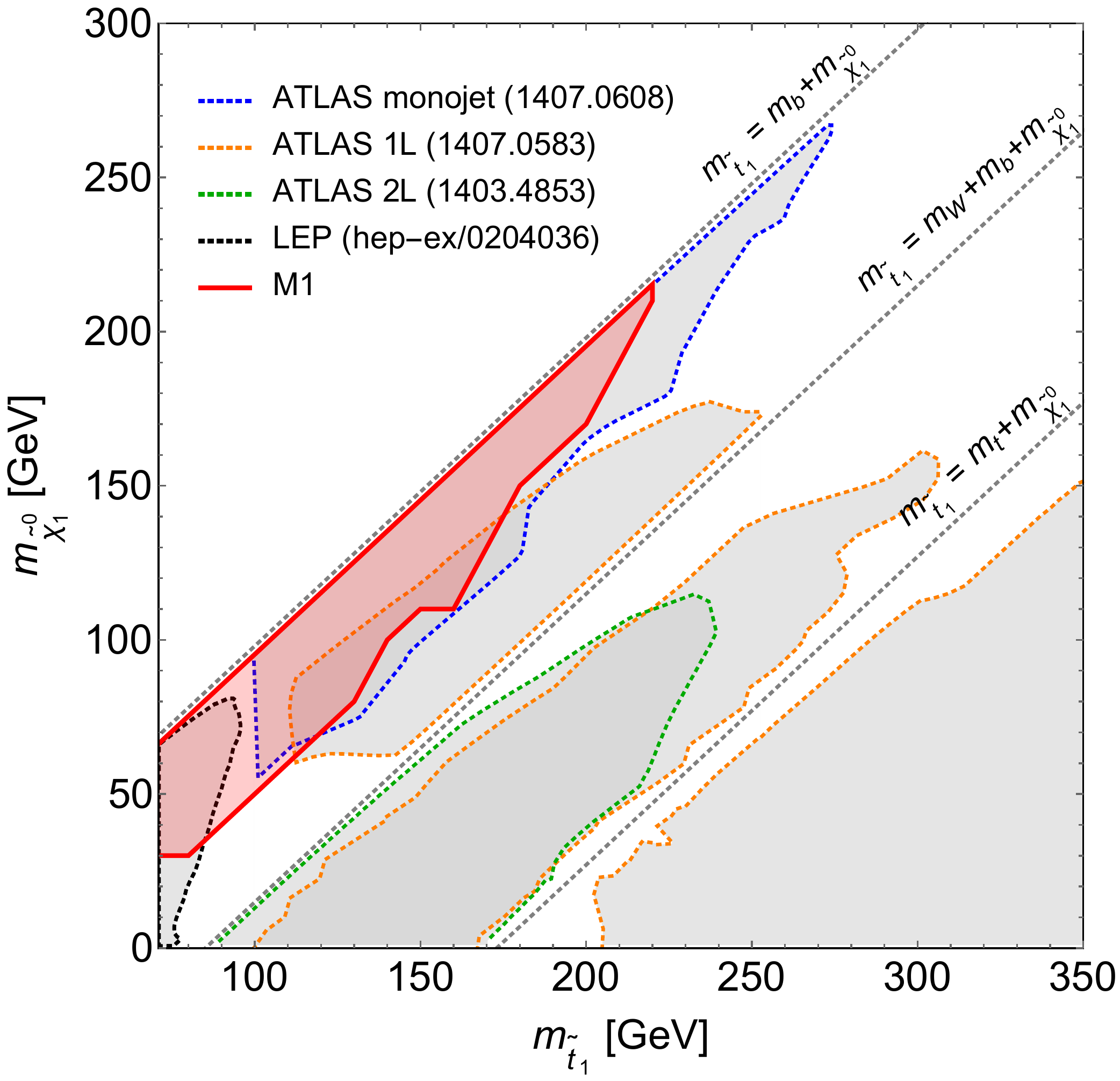}\hspace{8mm}
\hspace{-4.3mm} \includegraphics[scale=0.37]{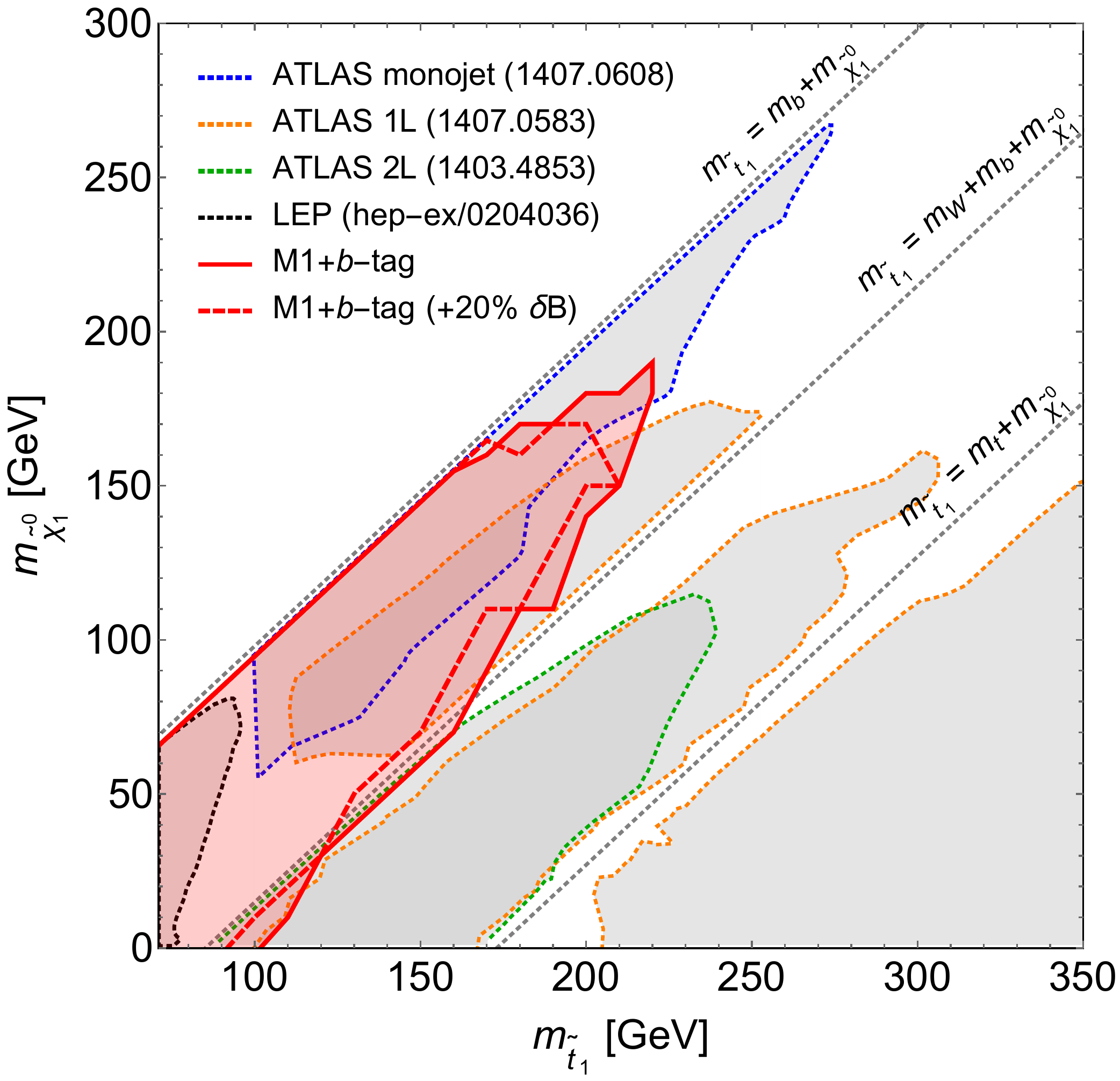}
\caption{Left panel: Summary of the existing limits in the stop-neutralino mass plane, as well as our exclusion curve for the M1 signal region of the ATLAS search \cite{ATLAS:2013aia}, in which we also extend the range of excluded stop masses below 100\,GeV. Right panel: Exclusion arising from our proposed search M1+$b$-tag, indicated by the red solid curve. The red dashed curve denotes the change caused by a 20\% increase of the total background error.}\vspace{-5.5mm}
\label{SummaryPlots}
\end{center}
\end{figure*}

The first search that places a limit in the stop 4-body decay case is a monojet search in which events are required to contain at least one hard jet and a large amount of missing transverse energy ($\MET$) \cite{ATLAS:2013aia}. 
The exclusion curve arising from this ATLAS search is indicated by the blue dashed curve in \mbox{Figure~\ref{SummaryPlots} (left).} As can be seen in the Figure, this search is most sensitive to the case where the mass splitting $\Delta m$ is small\footnote{For $\Delta m$ smaller than about 20\,GeV, the partial width for the stop 4-body decay decreases to the point where the stop either decays via a displaced vertex or, if other decay channels are present, the 4-body branching ratio is strongly suppressed.
However, in the spirit of simplified models, we follow the same strategy of ATLAS and present our results for 100\% 4-body decay branching ratio, assuming the stop to always decay promptly.
} and the stop \mbox{4-body} decay products are soft. The required hard jet arises from initial state radiation (ISR), against which the pair-produced stops recoil. The ISR jet boosts the two stops, which are no longer produced back-to-back, thereby increasing the $\MET$ in the event.

The abrupt end of the blue curve at $m_ {\tilde{t}_1}{=}100\,$GeV in Figure \ref{SummaryPlots} (left) is simply due to the fact that ATLAS does not provide the limit for smaller stop masses. Given the requirements in this search, one would expect that the exclusion curve should continuously extend diagonally down to the left, reaching the LEP limit \cite{Heister:2002hp}, which is indicated by the black dashed curve. In the Section Results below we discuss this issue further and provide the expected limits for stop masses below 100\,GeV.

The second ATLAS search that places a bound in the stop case under consideration is a search in the final state with one lepton, jets and $\MET$ \cite{Aad:2014kra}. Since this search relies on the presence of a lepton, arising from the stop 4-body decay, it is most sensitive to the case where the mass splitting $\Delta m$ is at least sufficiently large to allow for a reconstruction of the lepton. The exclusion curve arising from this search is indicated by the orange curve in \mbox{Figure~\ref{SummaryPlots} (left).} Note that this curve ends at neutralino and stop masses around 60\,GeV and  110\,GeV, respectively. Unlike the monojet search, in which the abrupt ending of the exclusion curve was due to the lack of signal samples, the reason for this ending has a physical origin. As one moves diagonally down to the left, i.e. keeping fixed $\Delta m$, the $\MET$ spectrum becomes softer and the search looses sensitivity \cite{Strandberg}.
Hence, while the cross-section increases, as a consequence of the decreasing stop mass, the acceptance times efficiency decreases faster, as a consequence of the decreasing neutralino mass. 

Figure~\ref{SummaryPlots} (left) summarizes the current experimental status concerning searches for light stops that dominantly decay in a 4-body final state. We see that there is a triangular shaped unconstrained region for stop masses in the range $90{-}140$\,GeV, with boundaries given by the exclusion curves from LEP \cite{Heister:2002hp}, the ATLAS monojet search \cite{ATLAS:2013aia} and the ATLAS \mbox{1-lepton} search \cite{Aad:2014kra}.

{\emph{Proposed stop 4-body search}.}---The search we propose is a simple extension of the ATLAS inclusive monojet search \cite{ATLAS:2013aia} where, in addition, we require the presence of at least one $b$-tagged jet. This $b$-tag requirement is motivated by how efficiently it reduces the leading background processes in Ref.~\cite{ATLAS:2013aia}, with respect to the stop 4-body decay signal process, thus enhancing the sensitivity to light compressed stops.

In order to display the potential gain in sensitivity that can be achieved, we consider the addition of a $b$-tag to the definition of the signal region ``M1" of the ATLAS monojet search \cite{ATLAS:2013aia}. The M1 selections require the presence of at most three jets (including $b$-jets)  with $p_T\,{>}\,30$\,GeV and $|\eta|\,{<}\,2.8$. The leading jet $p_T$ must be greater than 280\,GeV and the $\MET$ in the event must be above 220\,GeV. There is a veto on muons with $p_T\,{>}\,10$\,GeV and $|\eta|\,{<}\,2.4$ and on electrons  with $p_T\,{>}\,20$~GeV and $|\eta|\,{<}\,2.47$. Moreover, there is a condition requiring that the azimuthal angle $\Delta\phi$ between the $\MET$-vector and each of the jet \mbox{$p_T$-vectors} should be larger than $0.4$. The signal region that selects events according to these cuts is denoted by M1, in accordance with the ATLAS notation.

We remark that the selections that define the signal region M1 might not be optimal over the entire range of mass spectra that we consider. For instance, one could apply the same argument of adding $b$-tagging information to the other signal regions defined in Ref.~\cite{ATLAS:2013aia}, in which harder $\MET$ cuts are employed. Moreover, as the mass splitting between stop and neutralino varies, one could optimize further the number of jets that defines the signal region. This is done for instance in the Dark Matter search in Ref.~\cite{ATLAS-Collaboration:2014xia} where two signal regions with at least one \mbox{$b$-jet} and large $\MET$ are employed. However, we find the M1 selections sufficient to make the case for the addition of a $b$-tag that we propose and hence, in this work, we choose to concentrate on this signal region. The benefit of M1 is that it is the signal region in Ref.~\cite{ATLAS:2013aia} with the lowest $\MET$ cut and therefore it is the most promising for low mass stops. Furthermore this signal region is defined more inclusively on the number of jets than the signal regions of Ref.~\cite{ATLAS-Collaboration:2014xia}, which allows us to make more reliable predictions for the signal and background, and also to better reproduce the ATLAS results.

The background and signal simulations were performed using {\sc MadGraph5} 
\cite{Alwall:2014vb}, {\sc Pythia6} %
\cite{2006JHEP...05..026S}, \mbox{{\sc FastJet3}} \cite{Cacciari:2011rt,Cacciari:2006vn} and {\sc Delphes3} \cite{deFavereau:2013fe} with the ATLAS standard detector specification. Jets (including $b$-jets) are reconstructed using the anti-$k_t$ clustering algorithm \cite{Cacciari:2008hb} with a jet radius parameter of $0.4$. We have used the CTEQ6L1 PDF sets~\cite{Pumplin:2002vw} and MLM jet matching \cite{MLM1,Mangano:2006cp} throughout the analysis.

The leading SM backgrounds for our analysis are the $t\bar{t}$, $Z({\to}\nu\nu)+$jets, $W({\to}\ell\nu)+$jets (where $\ell\,{=}\,e,\mu,\tau$) and diboson processes. For the normalization of the cross sections of these processes at 8TeV LHC, we have used the same theoretical predictions as ATLAS in Ref.~\cite{ATLAS:2013aia}, obtained from Refs.~\cite{ATLAScollaboration:2012dh,Czakon:2011dq,Czakon:2013iy,Campbell:HraCaz2w,Campbell:2011jv,Catani:2007hl,Catani:2009hd}. We have validated our simulations against the M1 signal region of the ATLAS monojet search. Comparing our estimates for the background rates with the corresponding ATLAS estimates in Table VIII of Ref.~\cite{ATLAS:2013aia}, we find that the central values from our simulations are within 20\% of the corresponding ATLAS numbers. This agreement is quite satisfactory and makes us confident that reliable conclusions can be drawn using our tools.  To further improve our estimates, we normalize our predictions for all the leading background processes to exactly match ATLAS background estimates, which, for the $Z+$jets and $W+$jets samples, have the extra advantage of incorporating data-driven re-weighting of the simulated events. The expected backgrounds are summarized in Table~\ref{table:BGevents} written as $B\pm\delta B$, where $B$ is the central value and $\delta B$ the $1\sigma$ error. For all the backgrounds shown in Table \ref{table:BGevents} we quote 1$\sigma$ errors from ATLAS.

The crucial ingredient in our analysis is that we extend the M1 set of selections by adding the requirement that the events passing the M1 cuts must contain at least one $b$-tagged jet with $p_T$ in the range $30{-}300$\,GeV, and with $|\eta|\,{<}\,2.5$. The  $b$-tagging is parametrized  through {\sc Delphes3}~\cite{deFavereau:2013fe} using a ``mild'' working point characterized by a light flavor jet rejection of 1/1000 and an efficiency for actual $b$-quarks of around 0.4 for central jets. We emphasize that for the $p_T$ range that we have chosen for the $b$-tagged jet, the calibration of the $b$-tagging algorithms in the ATLAS experiment is data-driven, minimizing the systematic uncertainties on $b$-tagging coming from Monte Carlo simulations \cite{Coccaro}.
We denote this signal region by ``M1+$b$-tag".

In the row M1+$b$-tag in Table~\ref{table:BGevents} we show the number of background events we get in the proposed signal region. The comparison of the events expected for the M1 and for the M1+$b$-tag signal regions gives an idea of the power of the $b$-tag requirement to enhance the sensitivity to the signal. In fact, the stop signal is expected to behave similarly to the $t\bar{t}$ background, hence to be only mildly reduced by the $b$-tag requirement.  Typical signal efficiencies to the $b$-tag requirement are around $20\%$. In contrast, we observe a great suppression of the $W$ and $Z$ boson backgrounds, which are the leading backgrounds in the M1 signal region. We stress that in our work the efficiency of signal and backgrounds to the $b$-tag requirement is given by the parametrization  implemented in {\sc Delphes3}~\cite{deFavereau:2013fe}, which is expected to be reliable. In order to draw conclusions on safe grounds, we take the relative error of the M1+$b$-tag background prediction to be {\emph{twice}} the relative error quoted by ATLAS for the M1 region. This gives background relative errors slightly smaller than those quoted by ATLAS~\cite{ATLAS:2013aia} in the signal regions involving $c$-tags. Given  that $b$-tags are under better  experimental control than $c$-tags, we expect our estimate of the uncertainties to be fair.

The signal process has been simulated for a grid of points with $m_{\tilde t_1}$ varying from 70\,GeV to 250\,GeV, and  $m_{\tilde \chi^0_1}$ varying from 0 to 200\,GeV, in steps of 10\,GeV. The only relevant SUSY production mode is stop pair production, with cross section given in Table~\ref{table:SignalXS}. For stop masses in the range 100--250\,GeV we have used the next-to-leading-order (NLO) plus next-to-leading-logarithm (NLL) stop cross sections used by \mbox{ATLAS} and given by the \mbox{LHC SUSY Cross Section Working Group \cite{LHCSUSY}}, to which an uncertainty of around 16\% is assigned. For the stop mass points below 100 GeV, which are not given by ATLAS, we computed the NLO cross sections using {\sc Prospino} \cite{Beenakker:1998fr} and normalized them with the available ATLAS cross sections for stop masses above 100\,GeV.  We compared the expected number of signal events we obtained to the corresponding numbers reported by \mbox{ATLAS} \cite{ATLAS:2013aia} in several points and we found a systematic overshoot of around $20\%$. Therefore we normalized the expected number of signal events by decreasing it of $20\%$ to match the ATLAS central values.

The efficiencies in the analysis are rather small, requiring the generation of a large number of Monte Carlo events. For the backgrounds in the M1 case we are able to generate a sufficient amount of fully jet-matched events to keep the statistical error below 10\%. However, for the signal, given the amount of points in the grid we aim to cover, we do not have the computer resources to generate fully matched events at a similar level of statistical uncertainty. We solve this problem by performing the analysis in two steps. Since all events that pass the cuts of our signal regions contain a hard jet, for each point of the grid, we generated both the exclusive zero-jet leading-order (LO) stop pair production process $p p \,{\to}\, \tilde{t}_1 \,\tilde{t}_1$ and the one-jet process $p p \,{\to}\, \tilde{t}_1\, \tilde{t}_1 j$, with jet $p_T\,{>}\,200$\,GeV. The ratio of these two cross sections is used to calculate the efficiency of the $p_T\,{>}\,200$\,GeV cut. The unmatched one-jet sample was then used for the grid in the analysis. We have checked the validity of this procedure in a few points  by generating fully matched inclusive samples, with sufficient statistics, and we found the two procedures to be in agreement within one statistical standard deviation.

{\emph{Results}}.---In order to estimate whether a point in the stop-neutralino mass plane is excluded, we compute the number of expected events for that point in a given signal region. We expect that the experiments can put a 95\%~C.L. exclusion  for those mass points that yield a number of signal events  greater than $N_{95}=1.96 \,\delta B$, where $\delta B$ is the total error in Table ~\ref{table:BGevents}.

Following this limit-setting procedure we start by calculating the exclusion we obtain using the M1 signal region. The resulting exclusion curve is given by the red solid curve in the left panel of Figure~\ref{SummaryPlots}.
We remark that our exclusion curve follows quite closely the one given by ATLAS, which further validates our procedure. It should be noted that our curve extends down to the LEP bound for $m_{\tilde{\chi}^0_1}\,{>}\,m_{\tilde{t}_1}{-}\,40\,$GeV, thereby covering part of the unconstrained region between the blue ATLAS curve and the black LEP curve. Hence, by considering stop masses below 100\,GeV, the existing ATLAS bound arising from the M1 signal region can be extended.

{\label{tx:redcurve} Following} the same limit-setting procedure for the signal region M1+$b$-tag, we obtain the red solid exclusion curve in the right panel of Figure~\ref{SummaryPlots}. This is the main result of this paper.
The dashed red curve corresponds to a 20\% increase of the total error on the background. The comparison of the two red curves gives an idea of the sensitivity of our result to $a$) the uncertainty associated with the signal cross sections, $b$) possible contributions from sub-leading backgrounds not evaluated for M1+$b$-tag. We see that our proposed search M1+$b$-tag covers the entire unconstrained region. Moreover, it slightly extends the existing LHC limits for stop mass around 200\,GeV.

\begin{table*}[t!]
\small\begin{center}
\begin{tabular}{l|ccccccccccccccccccc}
$m_{\tilde t}$ {[GeV]}			&\ $70$	&\ $80$	&\ $90$	&\ $100$	&\ $110$	&\ $120$	&\ $130$	&\ $140$	&\ $150$	&\ $160$	&\ $170$	 &\ $180$	&\ $190$	&\ $200$	&\ $210$	&\ $220$	&\ $230$	&\ $240$	&\ $250$	\\ \hline
$\sigma(\tilde t \bar{\tilde{t}})$ {[pb]} 	&\ $2797$	&\ $1550$		&\ $912$	&\ $560$	&\ $362$	&\ $240$	&\ $163$	&\ $113$	&\ $80.3$	 &\ $58.0$	&\ $42.6$	&\ $31.9$	&\ $24.2$	&\ $18.5$	&\ $14.3$	&\ $11.2$	&\ $8.78$	&\ $6.97$	&\ $5.58$	\\ \hline
\end{tabular}
\end{center}\vspace{-2mm}
\caption{\small\label{table:SignalXS} Inclusive stop pair production cross-sections used in the analysis.}\vspace{-4mm}
\end{table*}

It is worth mentioning %
that our simulations suggest that further sensitivity is gained by removing the $\Delta\phi$ condition, but keeping the same $b$-tag requirement as in M1+$b$-tag. In the M1 signal region, the $\Delta\phi$ condition is introduced to reduce the pure-QCD multijet background, for which the $\MET$ originates from jet mismeasurements. However, the $b$-tag requirement can be seen as an alternative to the $\Delta\phi$ cut since it is expected, already by itself, to dramatically reduce the multijet background. With our simulation tools, the estimation of the multijet background would not be reliable, therefore we do not attempt to estimate the gain in sensitivity to light stops that could be achieved by employing a looser $\Delta\phi$ cut. Instead we content ourselves with simply encouraging the experimental collaborations to also consider a signal region with a looser $\Delta\phi$ cut.



{\emph{Acknowledgments}.}---We would like to thank A.\,Coccaro, A.\,Djouadi, G.\,Polesello, S.\,Strandberg and A.\,Wulzer for discussions, and B.\,Fuks and B.\,Nachman for comments on the draft. The work of C.\,P.~is supported by the Swedish Research Council (VR) under the contract 637-2013-475, by IISN-Belgium (conventions 4.4511.06, 4.4505.86 and 4.4514.08) and by the ``Communaut\'e Fran\c{c}aise de Belgique" through the ARC program and by a ``Mandat d'Impulsion Scientifique" of the F.R.S.-FNRS.
The work of R.T. was supported by the ERC Advanced Grant no.~267985 {\it DaMeSyFla} and by the Italian PRIN no.~2010YJ2NYW$\_$003. R.T. thanks A.~Bersani, A.~Brunengo, M.~Gravino for computing support. We also acknowledge, for computing resources, Heidi, the computing support of INFN Genova and INFN Padova, and the grant SNF Sinergia no.~CRSII2-141847.



\bibliographystyle{mine}
\bibliography{bibliography}

\end{document}